\documentclass{article}

\usepackage{arxiv}

\usepackage[utf8]{inputenc} 
\usepackage[T1]{fontenc}    
\usepackage{hyperref}       
\usepackage{url}            
\usepackage{booktabs}       
\usepackage{amsfonts}       
\usepackage{nicefrac}       
\usepackage{microtype}      
\usepackage{lipsum}
\usepackage{graphicx}

\graphicspath{ {./Res/} }

\title{\LARGE \bf
A NEW ARCHITECTURE FOR HAND-WORN SIGN LANGUAGE TO SPEECH TRANSLATOR.
}

\author{
	Sai Charan Bodda\thanks{bodda.charan@iitg.ac.in, b.saicharan@samsung.com}\\
	Engineer\\
	Samsung Research Institute Noida\\
	\And
	Palki Gupta\thanks{palki.gupta3030@gmail.com}\\
	Engineer\\
	Samsung Research Institute Noida\\
	\And
	Gaurav Joshi\thanks{g.joshi@samsung.com}\\
	Engineer\\
	Samsung Research Institute Noida\\
	\And
	Ayush Chaturvedi\thanks{ayush.c@samsung.com}\\
	Staff Engineer\\
	Samsung Research Institute Noida\\
}

\begin{document}
	\maketitle
	
	\begin{abstract}
		People with speech and hearing impairments often rely on sign language to communicate with others but most of the general population cannot understand sign language and sign language itself is a difficult language to learn, so there is a definite need for technologies to translate sign language to speech. In this paper, we describe the design and implementation of Smart glove, a hand-worn hardware device capable of translating American Sign Language gestures into English speech by tracking the finger's orientation, gestures and hand motion. It uses hardware sensors like Flex, Accelerometer and gyroscope and intelligent software to capture and translate the gestures into speech. This paper explains the translation of both Alphabet and Word gestures. New approaches and algorithms are proposed and implemented to address hardware-dependent issues in existing glove based designs. The whole device is designed to be modular with distributed processing units to encourage modular enhancement, reducing complexity, and interrelation between subsystems.Decision Trees are used in gesture recognition and error correction. We hope that the henceforth mentioned design and architecture would be the basis for the advancement in research related to sensor-based sign language translation along with research for smart glove and cybernetic accessories.
		\\
	\end{abstract}


	\section{Introduction}
	
	People rely on languages and gestures for communicating with each other, and these languages vary depending on the region. Languages can be classified broadly into two types. Vocal-based languages and Gesture-based languages. Sign language is a form of Gesture-based language used often by Deaf and Mute people to communicate with the rest of the world. But there are very few people who can understand Sign Language. So there is a definite need for technology-aided Sign language translators.
	\\
	Sign language heavily relies on hand movements and orientation. So technology translators for translation of Sign language require a means to capture the hand movements and their orientation. Even though there is ongoing research in sign language translation since the late '80s. It is slow-paced because of its larger hardware dependency.  Sign language translation approaches can be broadly classified into two categories, Camera-based and Hand-worn sensor-based translators with each approach having its own set of both advantages and disadvantages.
	\\
	In Camera-based approach, a camera is used to record the hand movements and orientation. Computer vision techniques are employed on the above recordings for gesture detection. some of the drawbacks of this technique are, gestures should be made in the camera's field of view, the approach requires high real-time processing capabilities .
	\\
	Sensor-based technologies are often developed into a glove.Sensors like Flex, EMG and IMU's are embedded into the glove and are used to record movements of hand and fingers. Even though this technology addresses a few major problems in Camera-based technologies, the signer has to wear the glove while signing.
		
	\begin{figure}[ht]
		\centering
		\includegraphics{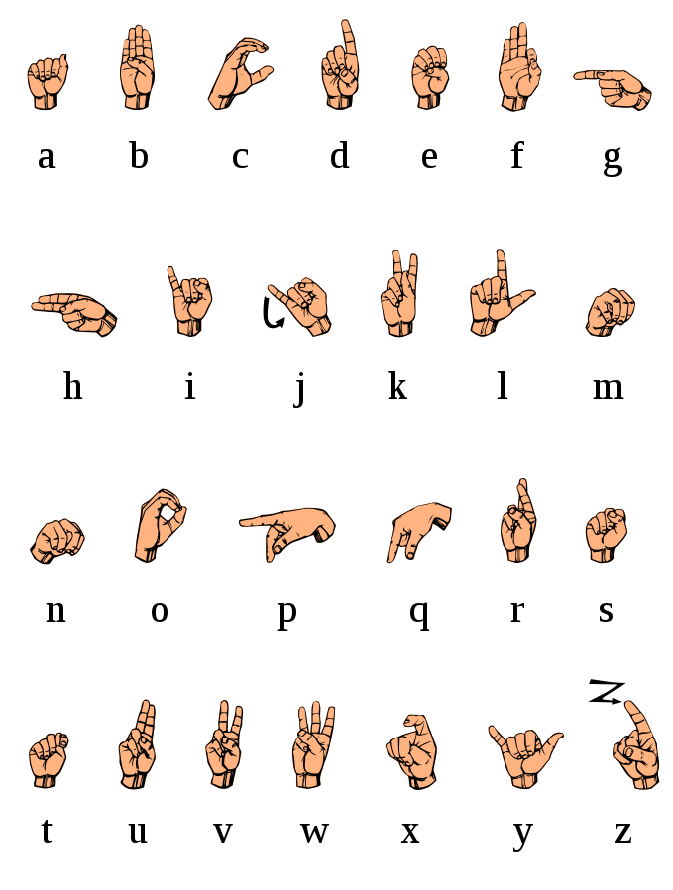}
		\caption{Gestures for Alphabets in American Sign Language\cite{c7}}
		\label{asl_gesture_images}
	\end{figure}
	
	The methodology section in the paper describes the architecture, design and implementation of a sensor-based sign language translator. The software and hardware are designed to be in a modular fashion to reduce the interdependency among another.
	\\
	
	Sign language gestures are composed of movements and orientation of hands and fingers. MPU 6050 is a MEMS-based Inertial Measurement Unit(IMU) with 6 degrees of freedom i.e., it has both 3- axis accelerometer and a gyroscope. MPU 6050 is used for hand orientation and movement tracking.  The Accelerometer and gyroscope values are used for hand tracking and determining hand orientation. Flex sensors are used for determining finger orientation. Flex sensors are variable resistors and their resistance value varies with the bending in the sensor. The processing requirement is quite high for real-time gesture recognition. So the data from sensors is encoded and sent to a companion smartphone device using a Bluetooth module. Ardunio nano is used for controlling the sensors ,and devices and for preliminary processing and encoding of the data. The data stream is encoded and custom trained Decision trees are used for gesture recognition.
	\\
	
		In this paper, we also propose techniques like automatic calibration of sensors because flex sensors are prone to have their resistance vary on continuous usage. The proposed approach can successfully recognize all alphabet gestures and can be trained to recognize word gestures. The glove is developed in a way for the signer to utter words either through word gestures or sign spelling out alphabets of the words. The proposed model can be  extended using Language models for word detection and auto-correction of misspelt alphabets.

	\section{Related Work}
	
	The work-related to Sign language translation can be broadly classified into two sections. Sensor-based sign language translation systems and Camera-based sign language translation systems. Sensor-based systems use IMU's for hand and finger tracking[\cite{c2,c4,h1}] and Camera-based systems record the user's hand movements and perform Computer vision algorithms for hand and finger tracking[\cite{c1,c3,c5}]. In \cite{c1} the authors used deep learning and computer vision to recognize American sign language. Their proposed model takes video sequences of sign making as input and extracts temporal (corresponding to hand movements) and spatial features (corresponding to hand position) from these video sequences.CNN is used for recognition of the extracted spacial features and RNN is used to train on the temporal features. Both NNs work together and recognize the gestures. since these models are trained on video sequences, there is visible training bias like dependency on skin tone. In \cite{c3} similar to \cite{c1} the authors used Camera-based systems, their model takes real-time video sequence or video recording of hand movements and uses optical-flow to interpret sign language. Optical flow works on determining how much an image pixel moves between two frame images sequentially in 2D. In this method, the user is required to wear a red glove for hand recognition and tracking across frames. In other researches like \cite{c5}, authors used HMM-based algorithms for gesture recognition. The major drawback in the above research is that the signer should perform signs in front of the camera and some times fingers or movements can be shadowed from the camera field of view. In \cite{c4} authors used hardware-based systems for sign detection. The authors used hall effect sensors and a strong magnet for finger tracking. A strong magnet is placed on the palm and hall sensors on the tip of the fingers. The value produced in sensors is related to distance from magnet and this information is used for finger position tracking. In \cite{c2} ,\cite{h1} , the authors used similar hardware sensors(flex, accelerometer and gyroscope) for getting finger position and orientation. The above methods \cite{c2,c4} ,\cite{h1} are hardware-dependent and glove and training process is custom for each person. Our research is an extension of \cite{h1} with methods to reduce hardware dependency issues and usage of ML rather than hard-coding for sensors.

	\section{Methodology}
	
		The product Smart Glove is designed to be a smartphone accessory with glove acting as a input device and processing is carried out in companion smartphone. we designed the smart glove to use the processing power of smartphone for processing heavy applications involved like gesture recognition and other machine learning algorithms.
	\\
	
	Most Sign language gestures are composed of fingers, hand and wrist movements and orientation. Flex sensors were used to measure the finger orientations and Accelerometer and Gyroscope are used to measure hand movements and orientation. These measured data is transmitted to a smartphone through the HC-05 Bluetooth Module. An application is developed in the smartphone to process the receive data, process the data and speaks out corresponding word or alphabet made.

	\subsection{Architecture and Algorithm Design}
	\begin{figure}[ht]
		\centering
		\framebox{\parbox{3in}{
				\includegraphics[scale=1]{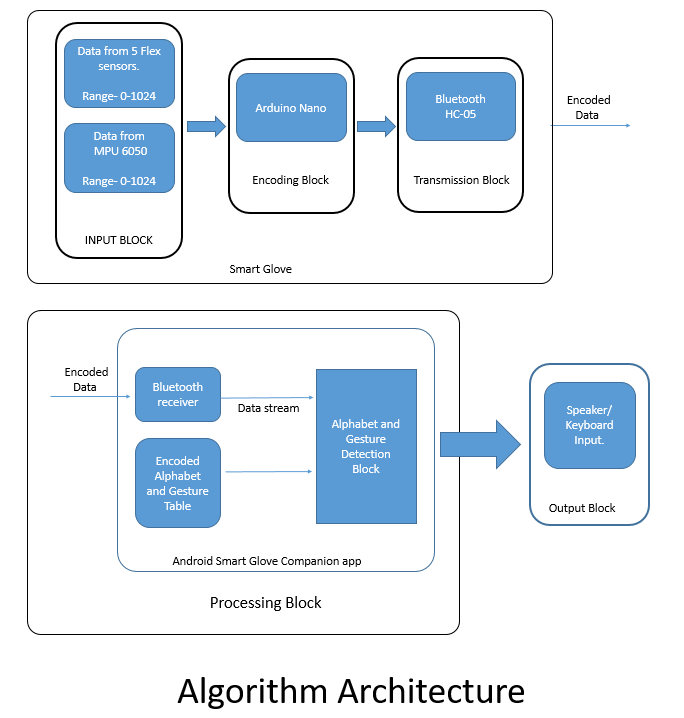}
		}}    
		\caption{System Architecture for Smart Glove}
		\label{figurelabel_architecture}
	\end{figure}
	
	On Each Individual Glove, 5 flex sensors, one for each finger, are used to measure finger orientation and bending of each finger and An 6-axis IMU Sensor(MPU 6050), with inbuilt Accelerometer and Gyroscope, is used for wrist and hand movements, orientation. The data from these sensors is sent to Encoding Block which is a microcontroller. In our case we used Ardunio Nano because of its compactness, this microprocessor quantizes the data from flex sensors into 3 states based on the ranges derived from Configuration phase(explained in later sections)  processed and converted to a stream of tertiary code for each sensor and send to the smartphone through Bluetooth device(HC-05) on the Glove. A custom App is built on the Smartphone which initializes Bluetooth receiver and collects these stream of data. the data is processed and corresponding Alphabet or word is interpreted and is spoken out of the smartphone speaker.
	
	\subsection{Hardware Details}
	
	As explained above, this whole Smart Glove can be broken down into two parts. Hardware and Processing part. Hardware part enables for capturing hand gestures and Processing Part deals with the interpretation of the gesture. Hardware part consists of Sensors, transmission device and Microcontroller for preliminary encoding and remaining component handling.
	.
	\subsubsection{Input Encoding and Transmission Blocks:}
	Flex Sensors are used for detection of finger state (i.e., how much finger is bent). As mentioned in Literature Review, Resistance of flex sensors increases with the increase in bending of the sensor. Flex sensors are connected in Voltage divider configuration with a resistor with a source voltage of 5V and voltage across the flex sensor is measured using Arduino nano analog pins.The Voltage divider is designed in a way that full bend flex sensor resistance is 2-5 times that of stright state. Arduino Nano's Analog input pins digitalize the voltage to 0-1024 with 0 corresponding to 0V and 1024 corresponding to 5V. More the bending in flex Sensor, the higher the resistance implying higher the voltage across flex sensors further implying higher values recorded by Arduino.
	\\
	
	Hand and wrist movements are an integral part of the sign language. these movements are measures using MPU 6050. MPU 6050 module is an Inertial Measurement Unit that has on accelerometer and gyroscope on a single chip. The accelerometer is used for finding Wrist orientation. Gyroscope is used for capturing wrist and hand movements as movements are part of most of every word gesture.  MPU 6050 uses I2C communication protocol for communication between devices. MPU 6050 has an inbuilt DSP and it covers sensors raw data into units g and dps for accelerometer and gyroscope. Considering Range and precision tradeoff for the sensor the output range is set to +2g to -2g for accelerometer and +500dps to -500 dps for gyroscope(dps=degree per second).
	\\
	
	The data from flex sensors and MPU are read periodically by Arduino Nano.The data is then encoded and transmitted to mobile phone via Bluetooth HC-05 sensor. The encoding process is explained in Software details section of this paper.
	\\
	
	After going through sign language gestures for alphabets, we concluded that it would be advantageous to quantize each finger position values into one of the 3 states, Straight, half bend and full bend. These quantized values from all finger along with values from MPU 6050 values are used for gesture recognition. The Process of quantization takes place in Encoding Block. The functionality of Encoding block, in our research, includes receiving data from sensors, quantizing the data and sending it to Bluetooth module for transmission. Arduino Nano microcontroller is used for controlling the sensors and encoding the data. It is made to run in loop infinitely ( reading the values from flex sensors and MPU 6050, encoding the data and transmits to the smartphone using Bluetooth. )
	\\
	
	This process of encoding is particularly useful and effectively address the major problem of repeatability with flex sensors as flex sensors value changes due to repeated use.

	\subsection{Software details}
	\subsubsection{Flowchart}
	
	\begin{figure}[ht]
		\centering
		\framebox{\parbox{6.2in}{\includegraphics[scale=2.1]{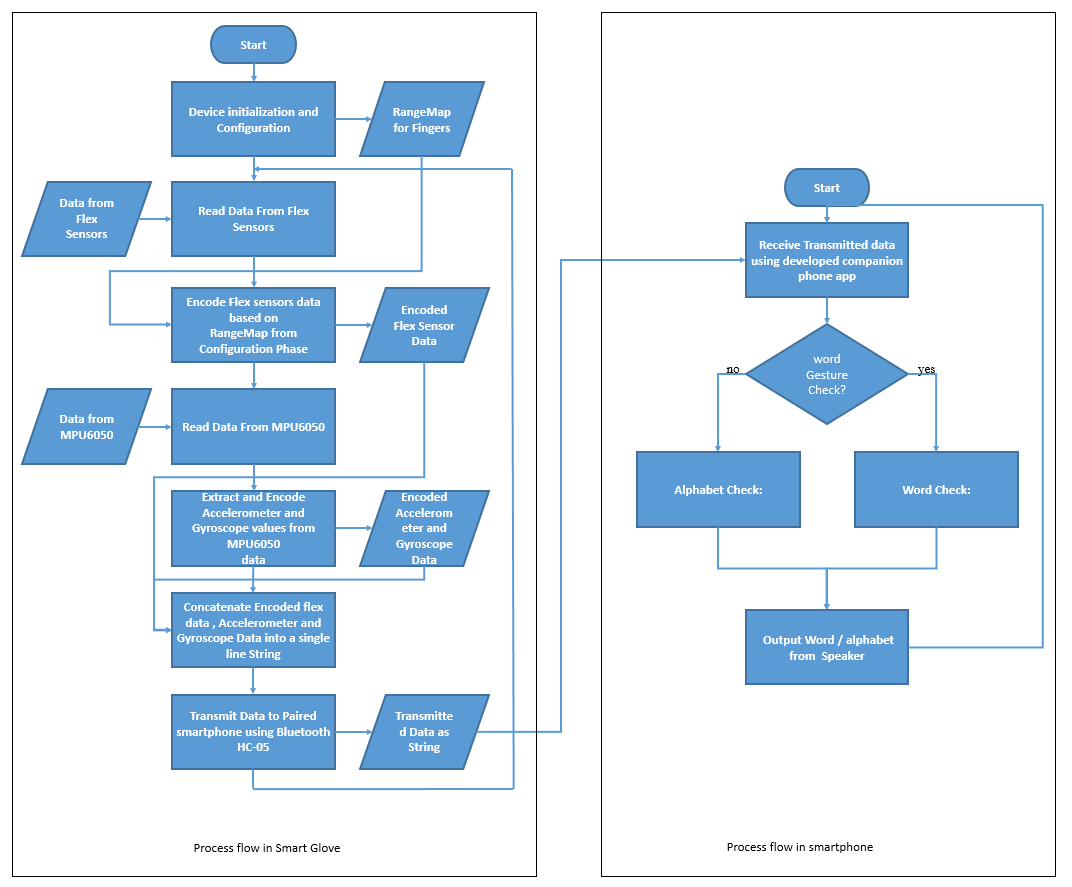}
		}}
		
		\caption{Software Program flow}
		\label{figurelabel_flowchart}
	\end{figure}
	
	The whole process mentioned in the flowchart can be divided into 2 parts with the processes on the left half of the flowchart are executed on the processor embedded in the glove and processes on the right half of the flowchart is executed on the companion smartphone. Both the processes run indefinitely in a loop. the finger and hand orientation and motion data are collected from sensors in the processes corresponding to the left half of the flowchart, processed and transmitted as a string to the second half of the loop. the second half of the loop processes the received data and performs gesture recognition algorithms and outputs the English translation of the made gesture.
	
	\subsubsection{Device Initialization and Configuration}
	
	This is the first stage the glove goes after every power on. Microcontroller, Sensors and Bluetooth are initialised and Bluetooth on the gloves scans and establishes communication with the Bluetooth of the companion phone. After the Initialisation stage, the device goes to the configuration phase. As it is established in the initial stages of our research, Flex Sensors are prone to produce different value for the same gesture over time. so hard coding ranges is not feasible. After every power-on, we programmed the device to go to the configuration phase and this phase lasts for about 10 seconds. During the phase, the user is required to make some random alphabet gestures, flex and close fists and fingers randomly. the microcontroller stores the stream of data from the flex sensors during this phase. this data is divided into 3 classes using K-means clustering. as we know that flex sensors produce the least value when they are straight (i.e., no bending) the class of values with least mean is considered as the Straight state. similarly, the class of values with the highest mean is considered as Full Bend state and class of value between straight and Full bend state are classified to be in Half Bend State.\\
	
	The above configuration phase makes sure that fully opened finger produces values in straight state and fully closed finger produces values in Full Bend State, and remaining values corresponds to Half Bend state. In the later sections of the paper, we explain that all possible finger states in ASL alphabets can be classified in one of the above states.
	\\
	
	The range of values produced by flex sensors when the finger is completely open (i.e., straight - little to no bending in flex sensor attached to finger ) is assigned as "Straight State" and is given value 1. The range of values produced by flex sensors when the finger is completely closed towards palm (i.e., tightly closed fist - maximum bending in flex sensors) is assigned as "Full bend State" and is given value 3. The range of between the "Straight State" and "Full bend State"(i.e., finger orientation between closed fist to completely open fist) is considered as "Half bend State" and is given value 2.
	
	\begin{figure}[ht]%
		\centering
		\framebox{\parbox{3in}{%
				\includegraphics[scale=1]{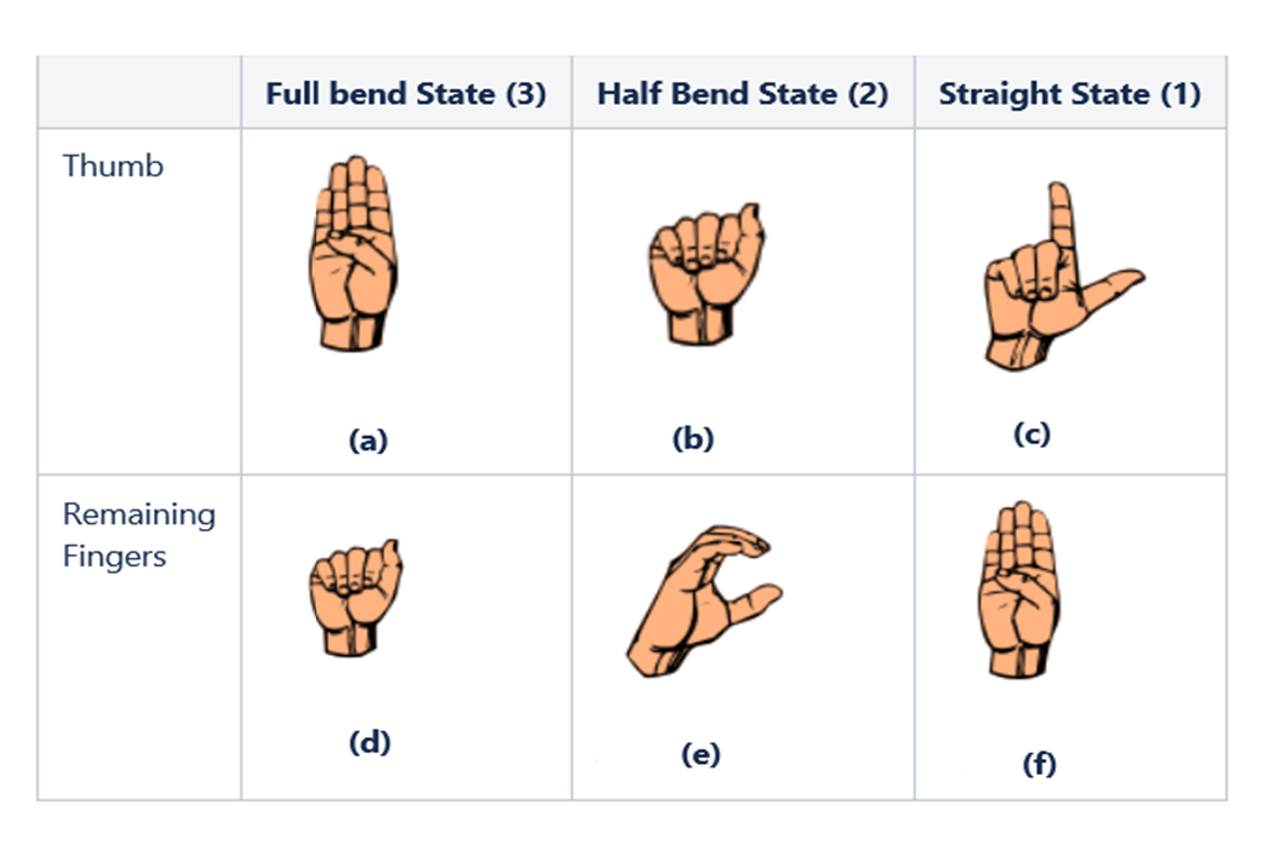}%
		}}%
		\caption{(a-c) : the range the values when thumb is approximately in position
			(a) is Full Bend ,(b) is half Bend ,(c) is Straight
			(d-f): the range the values when remaining fingers is approximately
			in position
			(d) is Full Bend ,(e) is half Bend ,(f) is Straight%
		}
		\label{Figurelabel_fingerstats}
	\end{figure}
	
	\begin{figure}[ht]
		\centering
		\framebox{\parbox{3in}{
				\includegraphics[scale=1]{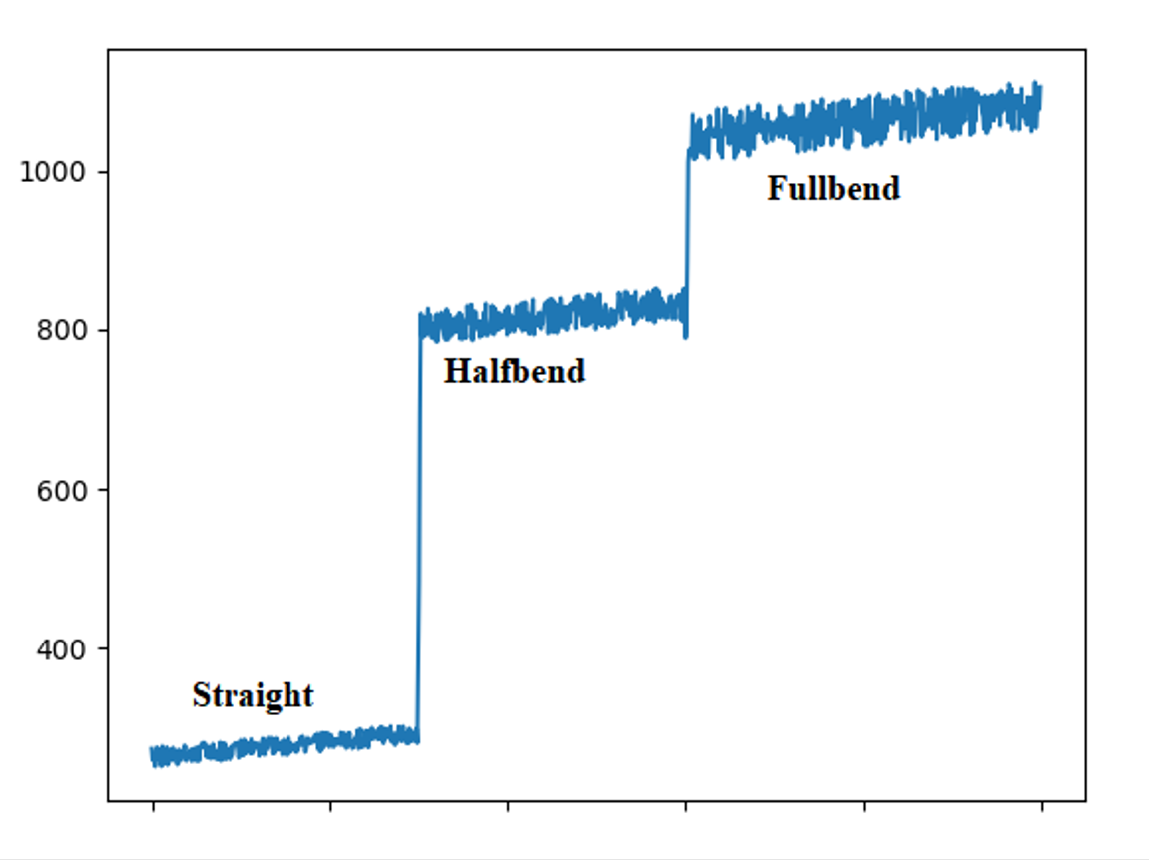}
		}}
		\caption{Flex Sensors values range when the fingers is Straight, Halfbend and Fullbend
		}
		\label{Figure_flexstates}
	\end{figure}
	
	Similarly, accelerometer values for alphabet gestures are quantized to 0 and 1 so gesture when the palm is vertical produces value 0 and when the palm is in the horizontal state produces 1.
	\\
	
	Since the gestures for alphabets, J and Z have movement for simplicity their gesture recognition is done similar to that of word gestures.

	\begin{table}
		\begin{center}
			\begin{tabular}{|c||c||c||c||c||c||c||c||c|}
				\hline
				[1] & [2]&[3]&[4]&[5]&[6]&[7] &[8]&[9]\\
				\hline
				A&2&3&3&3&3&0&0&233330\\
				\hline
				B&3&1&1&1&1&0&0&311110\\
				\hline
				C&1&2&2&2&2&0&0&122220\\
				\hline
				D&3&1&3&3&3&0&0&313330\\
				\hline
				E&3&3&3&3&3&0&0&322220\\
				\hline
				F&2&3&1&1&1&0&0&231110\\
				\hline
				G&2&1&3&3&3&1&0&213331\\
				\hline
				H&2&1&1&3&3&1&0&211331\\
				\hline
				I&3&3&3&3&1&0&0&333310\\
				\hline
				J&3&3&3&3&1&0&1&-\\
				\hline
				K&2&1&1&3&3&0&0&211330\\
				\hline
				L&1&1&3&3&3&0&0&113330\\
				\hline
				M&3&3&3&2&3&0&0&333230\\
				\hline
				N&3&3&2&3&3&0&0&332330\\
				\hline
				O&2&2&2&2&2&0&0&222220\\
				\hline
				P&1&1&2&3&3&1&0&112331\\
				\hline
				Q&1&2&3&3&3&1&0&123331\\
				\hline
				R&3&2&1&3&3&0&0&321330\\
				\hline
				S&3&3&3&3&3&0&0&333330\\
				\hline
				T&2&2&3&3&3&0&0&223330\\
				\hline
				U&3&1&1&3&3&0&0&311330\\
				\hline
				V&3&1&1&2&2&0&0&311330\\
				\hline
				W&3&1&1&1&3&0&0&311130\\
				\hline
				X&3&2&3&3&3&0&0&323330\\
				\hline
				Y&1&3&3&3&1&0&0&133310\\
				\hline
				Z&3&1&3&3&3&0&1&-\\
				\hline
				
			\end{tabular}
			
		\end{center}
		
		\caption{Table Corresponding to alphabet Map}
		\label{Alphabet_Map}
		
		\begin{center}
			\begin{tabular}{|c||c||c||c||c||c||c||c||c|}
				\hline
				[1] & [2]&[3]&[4]&[5]&[6]&[7] &[8]&[9]\\
				\hline
				U&3&1&1&2&2&0&0&311220\\
				\hline
			\end{tabular}
		\end{center}
		\caption{Updated Table Corresponding to Alphabet Map}
		\label{Table2}
		
		[1]: Alphabet, [2]: Thumb, [3]:    Index, [4]:    Middle, [5]: Ring, [6]:    Little, [7]: Wrist orientation    (Accelerometer), [8]: Hand Movement, [9]: Final Alphabet Code
	\end{table}	
	
	From TABLE I, all the code for alphabets are unique except for symbols (U, V). It is evident from the gestures chart that both Alphabets result in the same gesture code. To handle these collisions, the gesture for Alphabet U is slightly tweaked without modifying the gestures visual appearance. i.e., While making gesture U, The user needs to slightly relax ring and little finger so that they are interpreted as half bend. The above problem can be handled using Language models for words and error correction block.
	
	The generated Alphabet Code is stored in a Map data-structure and is used in Alphabet gesture recognition part.
	
	\subsubsection{Gesture recognition}
	
	The future developments in gesture recognition part can computation heavy based on the machine learning model and approach used, we designed a smart glove to be used as an lite weight accessory for smartphone. The companion smartphone can be used for processing and as the output device. After the configuration phase in glove, data from flex sensors and MPU 6050 is constantly encoded and transmitted to the smartphone every 50ms in the form of ASCII sting. An application is developed for the smartphone to receive data from the glove. This app extracts the data from the string and saves it in a buffer. The buffer maintains the data of 30 cycles. Statistical models like mean and variance are computed on the buffer data in every cycle and this information is used for gesture detection.
	\\
	
	If the computed variance values of accelerometer and gyroscope vary in every cycle and are high. The input gesture is considered as a word gesture and Word gesture recognition part is invoked . if the mean is constant and variance of the buffer data is low. The input gesture is considered as Alphabet and the corresponding process is invoked.
	
	\paragraph{Alphabet gesture recognition}
	During the course of gestures for alphabets (except J and Z whose gesture recognition is explained in Gesture recognition for words section) the hand and fingers are static and in a fixed position. So the accelerometer values don't vary much and the whole string of the data received by the smartphone is constant with little noise added. Since for alphabet gesture recognition, the only orientation of hand and fingers is required. The encoded part corresponding to flex and accelerometer are only considered. 
	\\
	
	To overcome noise effects and to avoid repeated multiple-output for a single uttered gesture. Buffer data of 1.5 sec is considered. Statistical Mode of the data is considered. The pattern corresponding to the above Mode is checked from the Alphabet code map. If a pattern match is found, the alphabet corresponding to matched pattern is sent to "Stream handling and Error Correction Block" and is outputted through phone speaker else the buffer is discarded and the same process is carried out for next stream of data.
	\\
	
	Stream Handling: Since there is a possibility that a single alphabet made over a long time produces multiple instances as output. to overcome this problem if the variance of data doesn’t go up between 2 consecutive gestures it is considered as a single gesture and outputted only once.
	
	\subsubsection{Word Gesture training and recognition}
	
	All Alphabet gestures (except for J and Z) are unique fixed and doesn't have spatial or temporal movements and produces steady accelerometer and gyroscope value. Most of the gestures corresponding to words include hand movement along with changes in finger orientation. i.e., the gesture includes variation across both temporal and spatial dimensions. Alphabets J and Z also have movements in their gesture. Their recognition is carried similar to other word gestures.
	\\
	
	For POC demonstration 4 Common word gestures hello, sorry, thank you and goodbye along with alphabets J and Z are considered. Gestures for alphabets J and Z along with above-mentioned words are referred to as Word gestures in this paper. Decision trees with 11 features(data point) (5 flex encoded values, 3 accelerometer values and 3 gyroscope values) are used for classification. Decision trees are trained with thousands of in-house labelled sample data. Each data sample is a 2sec long buffer of the first-order derivative of a data stream from the glove. The decision tree is trained to classify data into 7 classes (6 above-defined words and none if none of the words matches).The same approach can be carried out to include more and more gestures.
	\\
	
	The first-order derivative of the data stream is computed and stored in a temporary buffer. The data buffer is broken down into sample windows of 1.5 second long frames with 0.75-second long frameshift. the samples are fed to the decision tree and the output is spoken out. 
	
	\section{Conclusions and Future Work}
	
	This paper explains the architecture and design of the smart glove and hovers around the encoding of the spatial and temporal gestures made by hands into English alphabets and words. 
	\\
	
	The Design is made modular so that each of the sensors or devices can be upgraded without having a larger impact on other modules or software. The data from the sensors are encoded and the encoded data is used for the Machine learning calculations, processing and training. The Process of recognition of word gestures using decision trees is explored. the software implementation explained in this paper can be used for any sensors(other than flex) if the sensed data can somehow be quantized into 3 states thus making the whole software recognition part independent of hardware.
	\\
	
	The Future scope of this paper includes using language models for spelling out words using alphabet gestures. Error correction and word recognition based on context and previous words while spelling out words. using EMG sensors for finding finger orientation by measuring Muscle potentials so that the whole device can be made as a band and above explained architecture and software can be used.ANNs can be used in place of Decision trees for word gesture recognition to produce better results but requires a lot of training data. The device can be made to learn new gestures that can be used for controlling devices or in AR/VR games.
	
	\section{Conclusions and Future Work}
	
	This paper explains the architecture and design of the smart glove and hovers around the encoding of the spatial and temporal gestures made by hands into English alphabets and words. 
	\\
	
	The Design is made modular so that each of the sensors or devices can be upgraded without having a larger impact on other modules or software. The data from the sensors are encoded and the encoded data is used for the Machine learning calculations, processing and training. The Process of recognition of word gestures using decision trees is explored. the software implementation explained in this paper can be used for any sensors(other than flex) if the sensed data can somehow be quantized into 3 states thus making the whole software recognition part independent of hardware.
	\\
	The Future scope of this paper includes using language models for spelling out words using alphabet gestures. Error correction and word recognition based on context and previous words while spelling out words. using EMG sensors for finding finger orientation by measuring Muscle potentials so that the whole device can be made as a band and above explained architecture and software can be used.ANNs can be used in place of Decision trees for word gesture recognition to produce better results but requires a lot of training data. The device can be made to learn new gestures that can be used for controlling devices or in AR/VR games.

	\section*{ACKNOWLEDGMENT}
	
	We would like to acknowledge Mr Ankit Agarwal for guidance, suggestions, and for providing us with the opportunity to carry on research. We would also like to express our sincere gratitude to Mr Desh Deepak Agarwal and other co-workers in team Samsung Pass and Account for helpful suggestions and discussions.
	\\
	
	We would also like to thank Samsung Research Institute-Noida for providing computational and laboratory Resources.

	\bibliographystyle{unsrt}  


\end{document}